\documentclass[aip,reprint]{revtex4-1}

\usepackage{amsmath}
\usepackage{mathtools}
\usepackage{amssymb}
\usepackage{mathrsfs}
\usepackage{dsfont}
\usepackage{MnSymbol}
\usepackage{graphicx}%
\usepackage{enumerate} 
\usepackage{xcolor} 
\definecolor{a1}{rgb}{0,0,0.8}   
\usepackage{colortbl}	           
\usepackage[colorlinks=true,linktocpage=true,hyperindex=true,pageanchor=true,bookmarks=true]{hyperref} 
\hypersetup{citecolor=a1,linkcolor=a1,menucolor=a1,urlcolor=a1,runcolor=a1}
\hypersetup{citebordercolor=a1,linkbordercolor=a1,urlbordercolor=a1,runbordercolor=a1}
\hypersetup{hypertexnames=false} 

\newtheorem{theorem}{Theorem}%
	
\begin{document}


\title{Solutions of the Wheeler-Feynman equations with discontinuous velocities} 

\author{Daniel C\^{a}mara de Souza}
\email[author's email address:\ ]{daniel.souza@usp.br}
\affiliation{Universidade de S\~{a}o Paulo\\ Instituto de F\'{i}sica\\ Departamento de F\'{i}sica Matem\'{a}tica\\ Rua do Mat\~{a}o, Travessa R, N\,$^{\underline{o}}$ 187\\ Caixa Postal 66.318, CEP 05508-090 \\ S\~{a}o Paulo, S\~{a}o Paulo, Brasil}

\author{Jayme De Luca}
\email[author's email address:\ ]{jayme.deluca@gmail.com}
\homepage[webpage \\]{https://sites.google.com/site/jaymedeluca/}
\affiliation{Universidade Federal de S\~{a}o Carlos,\\ Departamento de F\'{\i}sica\\ Rodovia Washington Luis, km 235\\ Caixa Postal 676, CEP 13565-905\\ S\~{a}o Carlos, S\~{a}o Paulo, Brazil}

\date{\today}

\begin{abstract}
We generalize Wheeler-Feynman electrodynamics with a variational boundary-value problem with past and future boundary segments that can include velocity discontinuity points. Critical-point trajectories must satisfy the Euler-Lagrange equations of the action functional, which are neutral-differential delay equations of motion (the Wheeler-Feynman equations of motion). At velocity discontinuity points, critical-point orbits must satisfy the Weierstrass-Erdmann conditions of continuity of partial momenta and partial energies. We study a special class of boundary data having the shortest time-separation between boundary segments, for which case the Wheeler-Feynman equations reduce to a two-point boundary problem for an ordinary differential equation. For this simple case we prove that the extended variational problem has solutions with discontinuous velocities. We construct a numerical method to solve the Wheeler-Feynman equations together with the Weierstrass-Erdmann conditions and calculate some numerical orbits with discontinuous velocities.
\end{abstract}


\maketitle 

\section{Introduction}\label{SecIntrod}

\par To overcome the lack of equations of motion for point charges in Maxwell's electrodynamics, J. Wheeler and R. Feynman \cite{WheelerFeynman_1945, WheelerFeynman_1949} developed an electrodynamics based on the minimization of the Fokker-Schwarzschild-Tetrode \cite{Fokker_1929, Schwarzschild_1903, Tetrode_1922} action to replace Maxwell's equations. Besides having sensible equations of motion for point charges, the action functional seemed to point towards a canonical quantization\cite{WKB, HansWKB} of the two-body problem, so far an unfulfilled promise\cite{Mehra}.

\par In a recent development, Wheeler-Feynman electrodynamics was embedded in a variational boundary-value problem \cite{JMP2009}, henceforth variational electrodynamics\cite{minimizer,PIERB_2013}. The development of Ref. \cite{JMP2009} was followed by a study of the neutral differential delay equations of motion of the electromagnetic two-body problem \cite{minimizer,PIERB_2013}.

\par Consistent application of variational electrodynamics defines trajectories with discontinuous velocities among the critical points of the variational problem \cite{minimizer,PIERB_2013}, henceforth broken extrema \cite{Gelfand_1963}. The conditions for a piecewise $C^{2}$ extremal trajectory are (i) to satisfy the Euler-Lagrange equations along $C^2$ segments, which are neutral differential delay equations, henceforth the Wheeler-Feynman equations of motion, and (ii) at velocity discontinuity points (henceforth breaking points), a broken extremum must further satisfy the Weierstrass-Erdmann continuity conditions for the partial momenta and partial energies \cite{PIERB_2013, Gelfand_1963}, a system of four nonlinear equations at each breaking point \cite{Bellen_Zennaro_2003}.  It is known that neutral differential delay equations can generate breaking points \cite{Bellen_Zennaro_2003} which are propagated along trajectories \cite{Bellen_Guglielmi_2009}. As regards some physical property of the two-body problem,  it can be shown that globally bounded two-body trajectories with vanishing far-fields \emph{must} have discontinuous velocities \cite{minimizer}.

\par The electromagnetic variational problem \cite{JMP2009} has boundary conditions in past and future, which make it hard to study numerically. Here we study the simplest boundary value problem, having the shortest possible time-distance between past and future boundary segments, as explained in the following. For these, the variational problem reduces to a two-point boundary problem \cite{PIERB_2013} (i.e. a shooting problem  for an ODE integrator \cite{Ascher_Petzold_1998, *Ascher_1995}).  We show well-posedness and calculate some numerical solutions of the variational problem with shortest-length segments near circular orbits of large radii \cite{Schild_1963, Schonberg_1946}.

\par This paper is divided as follows: In Section \ref{SecVarProb} we introduce the variational boundary value problem. In Section \ref{SecODEandWECC} we explain the Weierstrass-Erdmann corner conditions for extrema with discontinuous velocities of the boundary-value problem with shortest-length boundary segments. In Section \ref{SecNum} we develop the numerical method for shortest-type boundaries and calculate some numerical trajectories. In Section \ref{SecDiscConc} we put the discussions and conclusion.

\section{Variational Boundary Value Problem}\label{SecVarProb}

\par Here we consider the natural electronic units where the speed of light, electronic charge and electronic mass are $c=-e_1=m_1 \equiv 1$. We henceforth use the index $i=1$ to indicate electronic quantities and the index $i=2$ to denote quantities of the positive charge $2$. 

\par The variational problem is to find (extended) trajectories given by continuous and piecewise $C^2$ functions of a (real) parameter $s \rightarrow  \boldsymbol{x}_{i}(s) \equiv (t_{i}(s),\mathbf{x}_{i}(s)) \in \mathbb{R} \times \mathbb{R}^3 \equiv \mathbb{R}^4 $. Henceforth a dot over vectors of $\mathbb{R}^4$  denotes derivative respect to the parameter s, i.e., $\boldsymbol{\dot{x}}_{i} \equiv d\boldsymbol{x}_{i}/ds=(\dot{t}_{i}(s),\mathbf{\dot{x}}_i(s))$ for $i=1,2$. On regular segments the particle velocities are recovered by the chain rule, $d\mathbf{x}_{i}/dt_i =\mathbf{\dot{x}}_i/\dot{t}_i$. The space $\mathbb{R}^4$ is equipped with the usual Euclidean norm of $\mathbb{R}^4$ (double bars), defined by the inner product, i.e., $\Vert\boldsymbol{x}_{i}\Vert^{2} \equiv \langle\boldsymbol{\dot{x}}_{i}|\boldsymbol{\dot{x}}_{i}\rangle \equiv t_{i}^{2}+|\mathbf{x}_{i}|^{2}$, where single bars $|.|$ denote the Euclidean norm in $\mathbb{R}^{3}$. To each vector $ \boldsymbol{v}=(a,\mathbf{b}) \in \mathbb{R} \times \mathbb{R}^3$ we define its dual by $\boldsymbol{v}^{\dagger}\equiv (a,-\mathbf{b}) $. A form that appears often is the (real) Minkowski bilinear form between two vectors $\boldsymbol{v}_1 \equiv (a_1,\mathbf{b}_1)$ and $\boldsymbol{v}_2 \equiv (a_2,\mathbf{b}_2)$, defined by the scalar product with the other vector's dual, i.e.,  $\langle \boldsymbol{v}_1|\boldsymbol{v}_2^{\dagger}\rangle \equiv a_1 b_2-(\mathbf{v}_1 \cdot \mathbf{v}_2)$, where dot denotes the scalar product of $\mathbb{R}^3$. 

\par Central to the construction of the variational problem are the light-cone conditions

\begin{equation}\label{lightcone_s1s2}
t_{j\pm}(s_{j})=t_{i}(s) \pm |\mathbf{x}_{i}(s)-\mathbf{x}_{j}(s_{j})|,
\end{equation}

\noindent for $j \equiv 3-i$ and $i=1,2$, which are implicit conditions for the trajectories. In Eq. (\ref{lightcone_s1s2}) the plus sign defines the future light-cone condition for particle i, while the minus sign defines the past light-cone condition for particle i. In Ref. \cite{JCAMRev7} it is shown that if trajectories are \emph{sub-luminal}, i.e.,

\begin{eqnarray}
\left\arrowvert\frac{d\mathbf{x}_{i}}{dt_i}\right\arrowvert<1,
\end{eqnarray}

\noindent for $i=1,2$, the light-cone conditions (\ref{lightcone_s1s2}) have unique solutions $s_{j\pm}(s)$. Henceforth a $\pm$ sign after the particle index $j \equiv 3-i$ indicates a quantity of particle $j$ evaluated at the advanced/retarded argument $s_{j\pm}(s)$.

\par The variational two-body problem \cite{JMP2009} is defined by the critical-point conditions of the action functional
 
\begin{eqnarray}\label{FTSJ}\!\!
S \!&\!\coloneqq\!\!&\!\!  \int_{s_{O^+}}^{s_{L_{B}}}\!\!\! K_{2}ds_{2} \,\!+\!\! \int_{s_{O_{A}}}^{\!s_{\!L^{- }}}\!\!\!\! K_{1}ds_{1} \,\!+\!\! \int_{s_{O_{A}}}^{s_{L^{+}}}\!\!\!\! U_{12}^{-}ds_{1} \,\!+\!\! \int_{s_{O_{A}}}^{s_{L^{-}}}\!\!\!\! U_{12}^{+}ds_{1}\!\,,\!\!\! \notag \\[2mm]
\end{eqnarray}

\noindent where

\begin{equation}\label{Ki}
K_{i} \coloneqq m_{i}\left(1-\sqrt{\langle\boldsymbol{\dot{x}}_{i} | \boldsymbol{\dot{x}}_{i}^\dagger\rangle}\right),
\end{equation}

\noindent and the interaction energy \begin{equation}\label{Uij}
U_{ij}^{\pm}(\boldsymbol{x}_{i},\boldsymbol{x}_{j\pm},\boldsymbol{\dot{x}}_{i},\boldsymbol{\dot{x}}_{j\pm}) \coloneqq -\frac{e_{i}e_{j}\langle\boldsymbol{\dot{x}}_{i} | \boldsymbol{\dot{x}}_{j\pm}^\dagger\rangle}{2 |\langle \boldsymbol{x}_{i}-\boldsymbol{x}_{j\pm} | \boldsymbol{\dot{x}}_{j-}^\dagger\rangle |},
\end{equation}

\noindent with $i=1,2$ and $j \equiv3-i$, which has a non-zero denominator along non-collisional trajectories  \cite{JCAMRev7}. Henceforth we specify for the attractive problem by setting  $e_{i}e_{j}=-1$ in Eq. (\ref{Uij}). The first variation of (\ref{FTSJ}) naturally decomposes in a sum of two partial variations, $\delta S = \delta S_1 + \delta S_2$, as follows; (i) for variation $\delta S_1$ trajectory 1 varies for $s_1 \in [s_{O_A}, s_{L^-}]$ while trajectory 2 is kept fixed, and therefore the first term on the right-hand-side of (\ref{FTSJ}) is a constant term, $I_{1}$. The remaining three integrals are over $ds_1$, 

\begin{eqnarray}\label{FTSJ2}\!\!
S  &\coloneqq& \ I_{1} + \int_{s_{O_{A}}}^{s_{L^{-}}}\!\! \mathscr{L}_{1}(\mathbf{x}_{1},\mathbf{v}_{1},\mathbf{x}_{2},\mathbf{v}_{2}) \,ds_{1},
\end{eqnarray}

\noindent thus defining the partial Lagrangian $\mathscr{L}_1$ as the integrand, 
and (ii)  along variation $\delta S_2$ trajectory 2 varies for $s_2 \in [s_{O^+}, s_{L_B}]$ while trajectory 1 is kept fixed. To calculate variation $\delta S_2$ it is convenient to express functional (\ref{FTSJ2}) with three integrals over $ds_2$, plus a constant term $I_2$, as obtained using a change of variable on the last two integrals of Eq. (\ref{FTSJ}) to the other particle's parameter in light-cone condition (\ref{lightcone_s1s2}) \cite{PIERB_2013}. Since the problem thus defined is totally symmetric, bellow we explain the critical point conditions and equations of motion for particle $1$ only. Form (\ref{FTSJ2}) with its integral over partial Lagrangian $\mathscr{L}_1$ is used to calculate the partial variation $\delta S_1$ and the Euler-Lagrange equations of motion of particle 1.

\par
\begin{figure}[!htbp]
\centering
\includegraphics[width=0.475\textwidth]{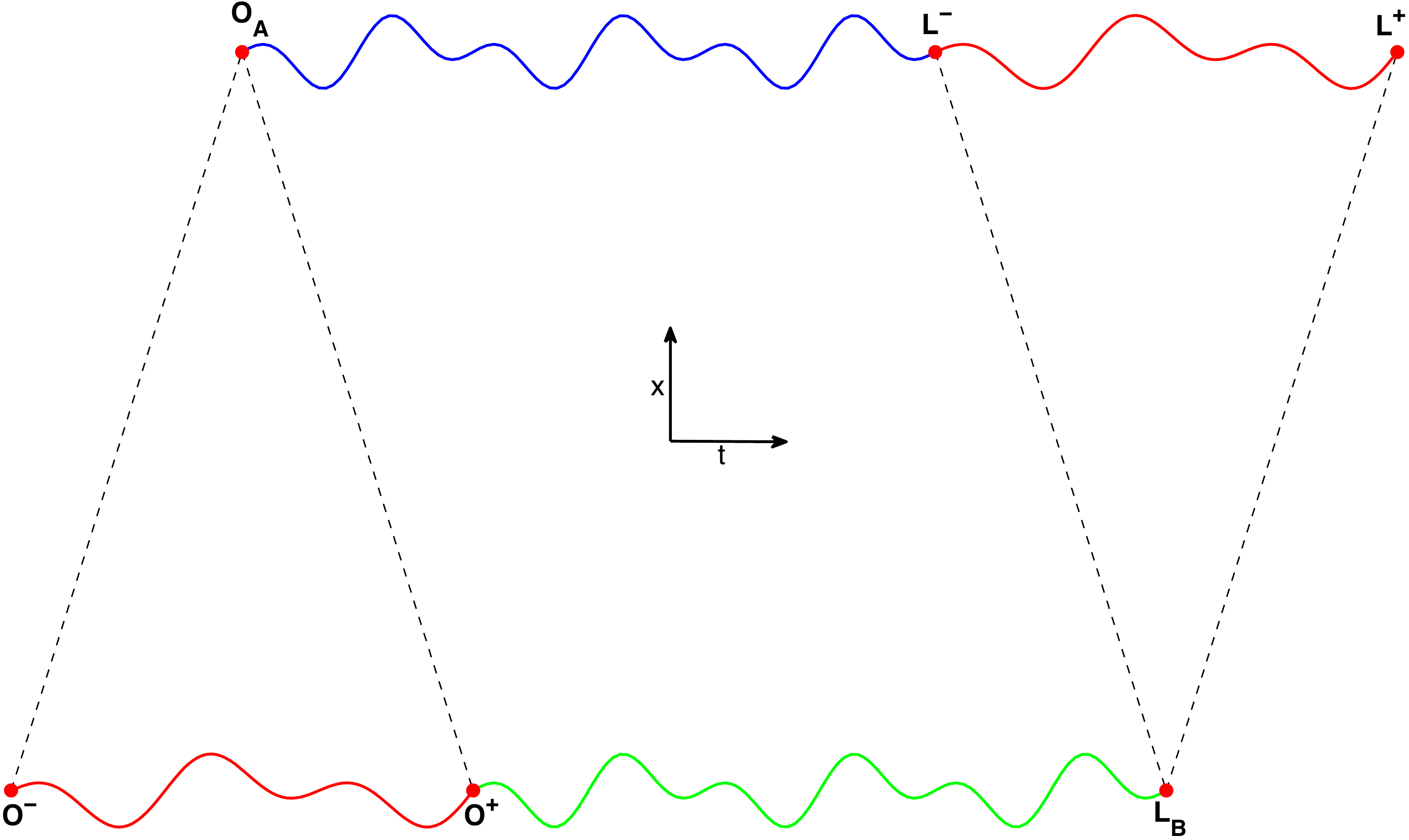}
\caption{\label{fig_boundary_value_problem} \footnotesize  The boundary conditions in $\mathbb{R} \times \mathbb{R}^{3}$ are (a)
initial point $ \boldsymbol{O}_{A} \equiv (s_{O_{A}},\mathbf{x}_1(s_{O_{A}}) )$ of trajectory  $1$
and the respective elsewhere boundary segment of $\mathbf{x}_2(s_2)$ for $ s_2\in [s_{O^{-}},s_{O^{+}}]$ (solid red line); (b)  endpoint $ \boldsymbol{L}_{B} \equiv (s_{L_{B}}, \mathbf{x}
_2(s_{L_{B}}))$ of trajectory $2$ and the respective elsewhere boundary segment of $%
\mathbf{x}_1(s_1)$ for $s_1\in[s_{L^{-}},s_{L^{+}}]$ (solid red line). Trajectories $\mathbf{x}_1(s_1)$
 for $s_1\in[s_{O_{A}},s_{L^{-}}]$ (solid blue line) and $\mathbf{x}_2(s_2)$ for $s_2\in[%
s_{O^{+}},s_{L_{B}}]$ (solid green line) are determined by the extremum condition.  Arbitrary units. }
\end{figure}

 The boundary conditions in past and future are described in Fig. \ref{fig_boundary_value_problem}, i.e. (a) the initial point $\boldsymbol{O}_{A}$ of
trajectory $1$ plus the boundary-segment of trajectory $2$ from point $\boldsymbol{O}^{-}$ in the past-light-cone of $\boldsymbol{O}_{A}$ up to point $\boldsymbol{O}^{+}$ in the future
light-cone of $\boldsymbol{O}_{A}$ (dashed black triangle on the left-hand-side of FIG. \ref{fig_boundary_value_problem}), and (b) the final point $\boldsymbol{L}_{B}$ of 
trajectory $2$ plus the boundary-segment of trajectory
$1$ inside the endpoints $(\boldsymbol{L}^{-}, \boldsymbol{L}^{+})$ in the past/future light-cone condition of $\boldsymbol{L}_{B}$ (dashed black triangle on the right-hand-side of FIG. \ref{fig_boundary_value_problem}).

\par Since past and future boundary segments are supposed to be independent of each other, we must have that the past boundary-segment of particle $2$ does not interact with the future boundary-segment of particle $1$, in which minimal case point $\boldsymbol{L}^{-}$ is in the forward light-cone condition of point $\boldsymbol{O}^{+}$.

\par Next we discuss the variational problem for piecewise $C^{2}$ continuous trajectories $\boldsymbol{x}_{1}(s)$, $\boldsymbol{x}_{2}(s)$ having monotonically increasing time-components and satisfying the above boundaries conditions. To calculate the first variation $\delta S_1$  we assume trajectory $\boldsymbol{x}_{i}(s)$ and its continuous perturbation $\mathbf{b}_{1}(s)$ are $C^2$ inside the intervals $s_{i}\in $ $(s_{\mu-1},s_{\mu})$ defined by the grid of possible discontinuities $s_\mu$ with $\mu=1,...,N$. The perturbed trajectory is defined by

\begin{equation}\label{perturb}
\boldsymbol{u}_{i}(\!s\!)\!=\!\boldsymbol{x}_{1}(\!s\!)+\mathbf{b}_{1}(\!s\!), 
\end{equation}

\noindent and outside the grid of discontinuity points, $s \neq s_\mu$,

\begin{equation}
\boldsymbol{\dot{u}}_{i}(\!s\!)\!=\!\boldsymbol{\dot{x}}_{i}(\!s\!)+\boldsymbol{\dot{b}}_{i}(\!s\!),
\end{equation}

\noindent and for $i=1$ satisfy the fixed-ends boundary conditions

\begin{equation} \label{boundaries}\!\!
\boldsymbol{b}_{1}(s_{O_{A}}\!)\!=\!\boldsymbol{b}_{1}(s_{L^{-}}\!)\!=\!0.
\end{equation}

\par The first variation $\delta S_i$ induced by a trajectory variation (\ref{perturb}) about a non-collisional sub-luminal trajectory is\cite{JCAMRev7}

\begin{eqnarray}\label{dsi}
\delta S_{i}&=&\sum\limits_{\mu=1}^{\mu=M}\int_{s_{\mu-1}}^{s_{\mu}}\left[\left.\left\langle\frac{\partial \mathscr{L}_{i} }{\partial \boldsymbol{x}_{i}}\right\arrowvert \boldsymbol{b}_{i}\right\rangle+\left.\left\langle\frac{\partial \mathscr{L}_{i} }{\partial \boldsymbol{v}_{i}}\right\arrowvert \boldsymbol{\dot{b}}_{i}\right\rangle\right]ds_{i}\notag \\[2mm]
&&+\ \mathcal{O}(\Vert \boldsymbol{b}_{i}\Vert ^{2}),
\end{eqnarray}

\noindent where $\Vert \boldsymbol{b}_{i}\Vert$ is the norm of piecewise $C^{2}$ perturbations,

\begin{equation}\label{normbiR4}
\Vert \boldsymbol{b}_{i}\Vert \coloneqq \sup\Vert\boldsymbol{b}_{i}\Vert + \operatorname{ess}\sup\Vert\boldsymbol{\dot{b}}_{i}\Vert + \operatorname{ess}\sup\Vert\boldsymbol{\ddot{b}}_{i}\Vert, 
\end{equation}

\noindent Integrating (\ref{dsi}) by parts over each $C^2$ segment to eliminate the integral containing $\boldsymbol{\dot{b}}_{i}(s)$ yields

\begin{eqnarray}\label{dsi2}
\delta S_{i} &\!=\!& \sum\limits_{\mu=1}^{\mu=N}\! \int_{s_{\mu-1}}^{s_{\mu}}\!\! \left\langle\boldsymbol{\boldsymbol{b}}_{i}\!\left\arrowvert \left[\frac{\partial \mathscr{L}_{i} }{\partial \boldsymbol{x}_{i}}-\frac{d}{ds}\!\left(\frac{\partial \mathscr{L}_{i} }{\partial \boldsymbol{v}_{i}}\right)\right]\right.\right\rangle ds\notag \\[2mm]  
&&+\ \sum\limits_{\mu=1}^{\mu=N}\! \int_{s_{\mu-1}}^{s_{\mu}} \!\!\frac{d}{ds}\left\langle\boldsymbol{b}_{i}(s) \left\arrowvert \frac{\partial \mathscr{L}_{i}}{\partial \boldsymbol{v}_{i}}\right.\right\rangle ds.
\end{eqnarray}

\noindent Since the $\boldsymbol{b}_{i}(s)$ are continuous and vanish at the endpoints, the second term of Eq. (\ref{dsi2}) can be re-arranged as

\begin{eqnarray}\label{arranje}
\delta S_{i} &\!=\!& \sum\limits_{\mu=1}^{\mu=N}\int_{s_{\mu-1}}^{s_{\mu}}\left\langle\boldsymbol{\boldsymbol{b}}_{i}(s)\left\arrowvert \left[ \frac{\partial \mathscr{L}_{i} }{\partial \boldsymbol{x}_{i}}-\frac{d}{ds}\left(\frac{\partial \mathscr{L}_{i} }{\partial \boldsymbol{v}_{i}}\right)\right]\right.\right\rangle ds
\notag \\[2mm] 
&&- \sum_{\mu=1}^{\mu=N-1}\left\langle \boldsymbol{b}_{i}(s_{\mu}) \left\arrowvert \Delta \left(\frac{\partial \mathscr{L}_{i} }{\partial \boldsymbol{v}_{i}}\right) \right.\right\rangle,
\end{eqnarray}

\noindent where

\begin{eqnarray}
\Delta \left(\frac{\partial \mathscr{L}_{i} }{\partial \boldsymbol{v}_i}\right) \coloneqq \left.\frac{\partial \mathscr{L}_{i} }{\partial \boldsymbol{v}_i}\right\arrowvert_{s_{\mu^+}} -\left.\frac{\partial \mathscr{L}_{i} }{\partial \boldsymbol{v}_i}\right\arrowvert_{s_{\mu^-}},
\end{eqnarray}

\noindent is the (possible) discontinuity of partial momentum $i$.

\par For a critical point we must have $\delta S_{i}=0$ for arbitrary $\boldsymbol{b}_{i}(s)$, and since the first term on the right-hand-side of (\ref{arranje}) is an integral and the second term depends on the discrete values of $\boldsymbol{b}_{i}(s_{\mu})$ on the finite number of grid points, each must vanish independently, yielding (a) Euler-Lagrange equations piecewise, henceforth the Wheeler-Feynman equations of motion, which can be written for the spatial components as \cite{JMP2009}

\begin{equation}\label{eqWF}
\!\!\!\!m_{i}\frac{d}{ds}\!\left(\!\frac{\mathbf{v}_{i}(s)}{\sqrt{1\!-\!\mathbf{v}_{i}^{2}(\!s\!)}}\!\right) \!=\! e_{i}[\mathbf{E}_{j}(\!s,\mathbf{x}_i(s)\!)+\mathbf{v}_{i}(s)\times\mathbf{B}_{j}(\!s, \mathbf{x}_i(s)\!)].\!\!\!\!
\end{equation}

\par As shown in Eq. (29) of Ref. \cite{JMP2009}, the fourth Euler-Lagrange equation (for the time component) vanishes identically if (\ref{eqWF}) holds. In Eq. (\ref{eqWF}), the $\mathbf{E}_{j}$ and $\mathbf{B}_{j}$ stand for the semi-sum of advanced and retarded Li\'{e}nard-Wiechert fields of charge $j$ \cite{Jackson_1999},

\begin{equation}
\mathbf{E}_{j} \equiv \frac{1}{2}\left(\mathbf{E}_{j}^{-}+\mathbf{E}_{j}^{+}\right), \qquad  \mathbf{B}_{j} \equiv \frac{1}{2}\left(\mathbf{B}_{j}^{-}+\mathbf{B}_{j}^{+}\right),
\end{equation}

\noindent where the Li\'{e}nard-Wiechert fields are defined by

\begin{equation}\label{EjBj}
\left\{\begin{array}{ll}
\!\mathbf{E}_{j}^{\pm}(s_{i},\mathbf{x}_{i}) \!\coloneqq \! \dfrac{\mathbf{u}_{ij\pm}(1-\mathbf{v}_{j\pm}^{2})}{\kappa_{ij\pm} r_{ij\pm}^{2}} \!+\! \dfrac{\mathbf{n}_{ij} \!\times\! \{\mathbf{u}_{ij\pm}\!\times\! \mathbf{\dot{v}_{j\pm}}\}}{\kappa_{ij\pm} r_{ij\pm}},\!\! \\[4mm]
\!\mathbf{B}_{j}^{\pm}(s_{i},\mathbf{x}_{i}) \! \coloneqq \! \mp \mathbf{n}_{ij\pm}\times \mathbf{E}_{j}^{\pm}.
\end{array}\right.
\end{equation}

\noindent with 

\begin{equation}\label{kappaij_uij}
\kappa_{ij\pm} \coloneqq 1\pm\mathbf{v}_{j\pm}\cdot\mathbf{n}_{ij\pm}, \qquad \mathbf{u}_{ij\pm} \coloneqq \mathbf{n}_{ij\pm} \pm \mathbf{v}_{j\pm}.
\end{equation}

\noindent Still in Eq. (\ref{EjBj}), $\gamma_{j\pm} \coloneqq (1-\mathbf{v}_{j\pm}^{2})^{-1/2}$ while $r_{ij\pm}$ is the distance in light-cone, $r_{ij\pm} \coloneqq |\mathbf{x}_{i}-\mathbf{x}_{j\pm}|$, and the unit vector $\mathbf{n}_{ij\pm} \coloneqq (\mathbf{x}_{i}-\mathbf{x}_{j\pm})/r_{ij\pm}$ points from the advanced/retarded position $\mathbf{x}_{j\pm}$ to the position $\mathbf{x}_{i}$. The vanishing of the second term on the right-hand-side of (\ref{arranje}) imposes four continuity conditions at each grid point, henceforth the Weierstrass-Erdmann corner conditions \cite{Gelfand_1963} of \emph{continuity} of partial momenta and partial energies

\begin{equation}\label{momento_energia_parcial}
\boldsymbol{P}_{i} \coloneqq \frac{\partial \mathscr{L}_{i} }{\partial \mathbf{v}_{i}}\quad\mbox{and}\quad
E_{i} \coloneqq - \frac{\partial \mathscr{L}_{i} }{\partial \dot{t}_{i}}.
\end{equation}

\noindent Using definition (\ref{momento_energia_parcial}) with Eq. (\ref{FTSJ}) yields the partial momentum

\begin{eqnarray}\label{momento_parcial}
\boldsymbol{P}_{i} & \coloneqq & \frac{m_{i}\mathbf{v}_{i}}{\sqrt{\dot{t}_{i}-\mathbf{v}_{i}^{2}}} - \frac{\mathbf{v}_{j-}}{2r_{ij-}(\dot{t}_{j-}-\mathbf{n}_{ij-}\cdot \mathbf{v}_{j-})} \notag \\[2mm]
&&\ - \frac{\mathbf{v}_{j+}}{2r_{ij+}(\dot{t}_{j+}+\mathbf{n}_{ij+}\cdot \mathbf{v}_{j+})},
\end{eqnarray}

\noindent and the partial energy

\begin{eqnarray}\label{energia_parcial}
E_{i} & \coloneqq & \frac{m_{i}\dot{t}_{i}}{\sqrt{\dot{t}_{i}-\mathbf{v}_{i}^{2}}} - \frac{\dot{t}_{j-}}{2r_{ij-}(\dot{t}_{j-}-\mathbf{n}_{ij-}\cdot \mathbf{v}_{j-})} \notag \\[2mm]
&&\ - \frac{\dot{t}_{j+}}{2r_{ij+}(\dot{t}_{j+}+\mathbf{n}_{ij+}\cdot \mathbf{v}_{j+})}.
\end{eqnarray}

\noindent Defining the right and left limits of the partial momenta/energies at each breaking point by $\mathbf{P}_{1}^{r}$, $\mathbf{P}_{1}^{l}$, $E_{2}^{r}$ and $E_{2}^{l}$, respectively, the Weierstrass-Erdmann corner conditions can be expressed at each grid point $s=s_\mu$ by

\begin{equation}\label{WECCsec2}
  \left\{
    \begin{array}{ll}
    	\Delta\mathbf{P}_{1} \coloneqq \mathbf{P}_{1}^{r} - \mathbf{P}_{1}^{l} = \mathbf{0}, \\[2mm]
	\Delta\mathbf{P}_{2} \coloneqq \mathbf{P}_{2}^{r} - \mathbf{P}_{2}^{l} = \mathbf{0}, \\[2mm]
    	\Delta E_{1} \coloneqq E_{1}^{r} - E_{1}^{l} = 0, \\[2mm]
	\Delta E_{2} \coloneqq E_{2}^{r} - E_{2}^{l} = 0.
    \end{array}
  \right.
\end{equation}

\section{Shooting Problem and Weiers\-trass-\-Erd\-mann corner conditions}\label{SecODEandWECC}

\par
\begin{figure}[!htbp]\centering
\includegraphics[width=0.475\textwidth]{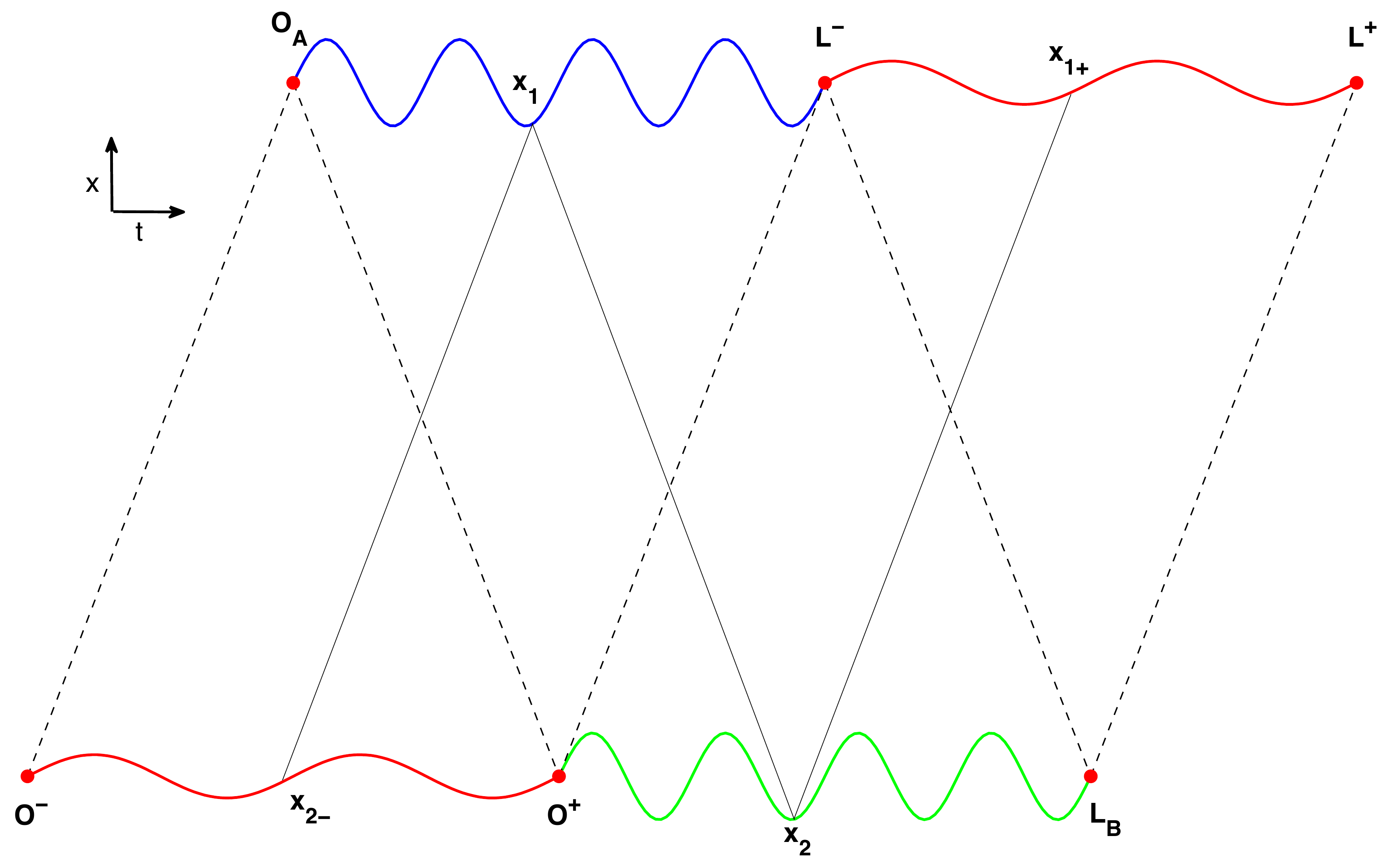}
\caption{\label{fig_bvp_short} Sketch of the space-time diagram for the boundary value problem with shortest-length boundaries. The variational problem is to find trajectories $\boldsymbol{x}_{1}(t)$, for $t\in[t_{O_{A}},t_{L^{-}}]$ (solid blue line), and $\boldsymbol{x}_{2}(t)$, for $t \in[t_{O^{+}},t_{L_{B}}]$ (solid green line), which match continuously with the boundary-segments $\boldsymbol{x}_{2}(t)$ with $t\in[t_{O^{-}},t_{O^{+}}]$ and $\boldsymbol{x}_{1}(t)$ with $t \in[t_{L^{-}},t_{L^{+}}]$. The event $\boldsymbol{O}^{+}$ is in the past light-cone of $\boldsymbol{L}^{-}$. Dashed and solid black lines connect points that are in the light-cone condition (either past or future).}
\end{figure}

 The shortest-length boundary value problem occurs when event $\boldsymbol{L}^{-}$ is in the future light-cone of event $\boldsymbol{O}^{+}$, as illustrated in Fig. \ref{fig_bvp_short}. Again, for boundaries with a smaller than the minimum time-separation illustrated in  Fig. \ref{fig_bvp_short}, the boundary-segments would interact in light-cone, an absurd. 

\par For shortest-length boundary conditions, points $\boldsymbol{x}_{2-}$ and $\boldsymbol{x}_{1+}$ fall each on a past/future boundary segment (illustrated in red in Fig. \ref{fig_bvp_short}), and therefore are given functions of the running positions $\boldsymbol{x}_{2}$ and $\boldsymbol{x}_{1}$, thus reducing the Wheeler-Feynman equations to a two-point boundary problem for an ODE, as explained in the following.

\par The Wheeler-Feynman equations of motion (\ref{eqWF}) can be expressed as \cite{PIERB_2013}

\begin{eqnarray}\label{eq_WF4}\!\!
m_{i}\gamma_{i} \mathbf{a}_{i} &=& \frac{e_{i}}{2}[(\mathbf{v}_{i}\cdot\mathbf{E}_{j}^{-})(\mathbf{n}_{ij-}-\mathbf{v}_{i}) + (1-\mathbf{v}_{i}\cdot\mathbf{n}_{ij-})\mathbf{E}_{j}^{-} \notag \\[2mm]
&-& (\mathbf{v}_{i}\cdot\mathbf{E}_{j}^{+})(\mathbf{n}_{ij+}+\mathbf{v}_{i}) + (1+\mathbf{v}_{i}\cdot\mathbf{n}_{ij+})\mathbf{E}_{j}^{+}],
\end{eqnarray}

\noindent where the fields $\mathbf{E}_{j}^{\pm}$ can be re-arranged using (\ref{kappaij_uij}) as

\begin{equation}\label{eq_Field2}
\mathbf{E}_{j}^{\pm}(t_{i},\mathbf{x}_{i}) \!=\! \dfrac{\mathbf{u}_{ij\pm}}{\gamma_{j\pm}^{2} \kappa_{ij\pm}^{3} r_{ij\pm}^{2}} \!+\! 
\dfrac{(\mathbf{n}_{ij\pm} \cdot \mathbf{a}_{j\pm})\mathbf{u}_{ij\pm}}{\kappa_{ij\pm}^{3} r_{ij\pm}} \!-\! \dfrac{\mathbf{a}_{j\pm}}{\kappa_{ij\pm}^{2} r_{ij\pm}}.\!\!
\end{equation}

\noindent The linear dependence of the $\mathbf{E}_{j}^{\pm}$ on the accelerations $\mathbf{a}_{j\pm}$ allows the use of a (simpler) matrix form

\begin{equation}\label{EDE}
\mathbf{E}_{j}^{\pm}(t_{i},\mathbf{x}_{i}) = \mathds{D}_{ij\pm} \mathbf{u}_{ij\pm}+\mathds{B}_{ij\pm}\mathbf{a}_{j\pm},
\end{equation}

\noindent where 

\begin{equation}\!\!
\mathds{D}_{ij\pm} \!\coloneqq\! \dfrac{e_{j}\mathds{1}}{\gamma_{j\pm}^{2} \kappa_{ij\pm}^{3} r_{ij\pm}^{2}}, \quad \mbox{and}\quad \mathds{B}_{ij\pm} \!\coloneqq\! \dfrac{e_{j}[\mathbf{u}_{ij\pm}\mathbf{n}_{ij\pm}^{\intercal} - \kappa_{ij\pm}\mathds{1}]}{\kappa_{ij\pm}^{3} r_{ij\pm}}.\!\!
\end{equation}

\par Further using (\ref{kappaij_uij}), Eq. (\ref{eq_WF4}) thus becomes

\begin{equation}\label{FKK}
m_{i}\gamma_{i}\mathbf{a}_{i} = \mathds{K}_{-ij}\mathbf{E}_{j}^{-} + \mathds{K}_{+ij}\mathbf{E}_{j}^{+},
\end{equation}

\noindent where

\begin{equation}
\mathds{K}_{\pm ij} \coloneqq \frac{e_{i}}{2}[ \mathbf{u}_{\pm ij}\mathbf{v}_{i}^{\intercal}+ \kappa_{\pm ij}\mathds{1}].
\end{equation}

\par We can isolate the accelerations $\mathbf{a}_{1}$ and $\mathbf{a}_{2}$ from Eqs. (\ref{eq_WF4}) with $i=1,2$, yielding an algebraic-differential equation

\begin{equation}\label{eqAlg}
  \left\{
    \begin{array}{ll}
    	(M_{1}M_{2}\mathds{1}-\mathds{A}_{12}^{+}\mathds{A}_{21}^{-}) \mathbf{a}_{1} = \mathds{A}_{12}^{+}\mathbf{F}_{1}^{+} + M_{2}\mathbf{F}_{2}^{-}, \\[2mm]
    	(M_{1}M_{2}\mathds{1}-\mathds{A}_{21}^{-}\mathds{A}_{12}^{+}) \mathbf{a}_{2} = \mathds{A}_{21}^{-}\mathbf{F}_{2}^{-} + M_{1}\mathbf{F}_{1}^{+},
    \end{array}
  \right.
\end{equation}

\noindent where we have defined the matrices

\begin{equation}
  \left\{\!
    \begin{array}{ll}
    	\mathbf{F}_{2}^{-} \!\coloneqq\! \mathds{K}_{-12}(\mathds{D}_{12-} \mathbf{u}_{12-} \!+\! \mathds{B}_{12-}\mathbf{a}_{2-}\!) \!+\! \mathds{K}_{+12}(\mathds{D}_{12+} \mathbf{u}_{12+}\!),\!\!\!\! \\[2mm]
    	\mathbf{F}_{1}^{+} \!\coloneqq\! \mathds{K}_{+21}(\mathds{D}_{21+} \mathbf{u}_{21+} \!+\! \mathds{B}_{21+}\mathbf{a}_{1+}\!) \!+\! \mathds{K}_{-21}(\mathds{D}_{21-} \mathbf{u}_{21-}\!),\!\!\!\!
    \end{array}
  \right.
\end{equation}
\noindent and $M_{i} \coloneqq m_{i}\gamma_{i}$, $\mathds{A}_{12}^{+} \coloneqq \mathds{K}_{+12}\mathds{B}_{12+}$ and $\mathds{A}_{21}^{-} \coloneqq \mathds{K}_{-21}\mathds{B}_{21-}$.

\noindent If one can invert the matrices on the left-hand-side of (\ref{eqAlg}) locally, the \emph{algebraic differential equation} (\ref{eqAlg}) reduces to an ODE.

\par In order to be able to invert the matrices on the left-hand-side of (\ref{eqAlg}), 

\begin{equation}\label{EqDifAlg}
  \left\{
    \begin{array}{ll}
    	\mathds{G}_{12} \coloneqq {(M_{1}M_{2}\mathds{1}-\mathds{A}_{12}^{+}\mathds{A}_{21}^{-})}, \\[2mm]
    	\mathds{G}_{21} \coloneqq {(M_{1}M_{2}\mathds{1}-\mathds{A}_{21}^{-}\mathds{A}_{12}^{+})},
    \end{array}
  \right.
\end{equation}
 
\noindent we restrict to boundary-segments satisfying

\begin{equation}\label{RelationInverseG1G2}
M_{1}M_{2}\gg \frac{1}{r_{12+}r_{12-}\kappa_{12+}^{3}\kappa_{12-}^{3}}.
\end{equation}

For example, along small perturbations of segments of circular orbits with large radii, $r_{ij\pm}\gg 1$ \cite{Schild_1963}, condition (\ref{RelationInverseG1G2}) holds.

\par Notice that the running accelerations in (\ref{eqAlg}) are each defined respect to a different independent time variable, i.e., 
$\mathbf{a}_1 \equiv d^2\mathbf{x}_1/d{t_1^2}$ and $\mathbf{a}_2 \equiv d^2\mathbf{x}_2/d{t_{2+}^{2}}$, where $t_{2+}(t_1)$ is defined by the implicit-function theorem and the future light-cone relation (\ref{lightcone_s1s2}). The derivative $dt_{2+}/dt_{1}$ is obtained by taking a derivative of (\ref{lightcone_s1s2}) piecewise, yielding

\begin{equation}
\lambda_{12+}  \coloneqq \frac{dt_{2+}}{dt_{1}} = \frac{(1+\mathbf{n}_{12+}\cdot\mathbf{v}_{1})}{(1+\mathbf{n}_{12+}\cdot\mathbf{v}_{2+})}.
\end{equation}

\noindent Last, if (\ref{RelationInverseG1G2}) holds we can define the vector fields

\begin{equation}
  \left\{
    \begin{array}{ll}
    	\mathbf{H}_{12} \coloneqq \mathds{G}_{12}^{-1} (\mathds{A}_{12}^{+}\mathbf{F}_{1}^{+} + M_{2}\mathbf{F}_{2}^{-}), \\[2mm]
    	\mathbf{H}_{21} \coloneqq \mathds{G}_{21}^{-1} (\mathds{A}_{21}^{-}\mathbf{F}_{2}^{-} + M_{1}\mathbf{F}_{1}^{+}),
    \end{array}
  \right.
\end{equation}

\noindent and transform (\ref{eqAlg}) into the following non-autonomous ODE

\begin{equation}\label{ET3}
  \left\{
    \begin{array}{ll}
    	\mathbf{\dot{v}}_{1} = \mathbf{H}_{12}(t_1,\mathbf{x}_{1},\mathbf{v}_{1},\mathbf{x}_{2},\mathbf{v}_{2}), \\[2mm]
    	\mathbf{\dot{v}}_{2} = \lambda_{12+}(t_1)\mathbf{H}_{21}(t_1,\mathbf{x}_{1},\mathbf{v}_{1},\mathbf{x}_{2},\mathbf{v}_{2}), \\[2mm]
	\mathbf{\dot{x}}_{1} = \mathbf{v}_{1}, \\[2mm]
	\mathbf{\dot{x}}_{2} = \lambda_{12+}(t_1)\mathbf{v}_{2},
    \end{array}
  \right.
\end{equation}

\noindent with the two-point boundary conditions

\begin{equation}\label{BC}
    \begin{array}{ll}
    	\mathbf{x}_{1}(t_{O_{A}})=\mathbf{x}_{{O}_{A}}, \qquad \mathbf{x}_{1}(t_{L^{-}})=\mathbf{x}_{{L}^{-}}, \\[2mm]
	\mathbf{x}_{2}(t_{O^{-}})=\mathbf{x}_{{O}^{-}}, \qquad \mathbf{x}_{2}(t_{L_{B}})=\mathbf{x}_{{L}_{B}},
    \end{array}
\end{equation}

\noindent thus defining a two-point boundary problem. The factor $\lambda_{12+} $ in Eq. (\ref{ET3}) insures the running positions satisfy the $t_1$-explict condition (\ref{lightcone_s1s2}), thus arriving at the end-point with $\mathbf{x}_{2}(t_{L_{B}})$ in the future light-cone of $\mathbf{x}_{1}(t_{L^{-}})$ (the second column of  boundary condition (\ref{BC})). 

\par We solve the two-point boundary problem (\ref{ET3}) and (\ref{BC}) with a shooting method that searches initial velocities $\mathbf{v}_{1}(t_{O_{A}})$, $\mathbf{v}_{2}(t_{O^{-}}^{r})$ at initial positions $\mathbf{x}_{{O}_{A}}$, $\mathbf{x}_{{O}^{-}}$ such that the initial value problem terminates at the specified end-points $\mathbf{x}_{{L}^{-}}$, $\mathbf{x}_{{L}_{B}}$. 

\par To define the map for the shooting method, we consider that position $\mathbf{X}\equiv [\mathbf{x}_{1},\mathbf{x}_{2}]^{\intercal}$ at time  $t_{{L}^{-}}$ depends on the initial velocity $\mathbf{V}\equiv [\mathbf{v}_{1},\mathbf{v}_{2}]^{\intercal}$, i.e.  $\mathbf{X}=\mathbf{X}(t_{{L}^{-}},\mathbf{V})$, and linearize about some reference initial velocity, $\mathbf{V}_{0}\equiv [\mathbf{v}_{10},\mathbf{v}_{20}]^{\intercal}$, yielding

\begin{equation}\label{map_shooting}
\mathbf{X} = \mathbf{X}_{0} + \mathds{J}_{s0}(\mathbf{V} - \mathbf{V}_{0}).
\end{equation}

\par Restricted to small perturbations of circular boundary segments, the velocity is a constant at $O(1/r_{12})$. The shooting method is used with (\ref{map_shooting}) to numerically calculate matrix $\mathds{J}_{s0}$ and vector $\mathbf{X}_{0}$ for the perturbed boundary data. For that we solve \emph{seven} initial value problems (\ref{ET3}) for the shooting map (\ref{map_shooting}). The seven initial velocities $\mathbf{V}$ are used with the same small perturbation of boundary segments from a large radius circular orbit,  defined as follows; (a) we start with the velocity of the unperturbed circular orbit, for which the second term on the right-hand-side of (\ref{map_shooting}) vanishes, thus calculating the constant $\mathbf{X}_{0}$ within the numerical precision, and (b) we perturb each of the six components of the initial velocity away from the circular orbit's velocity $\mathbf{V}_{0}$, one component at a time. We further solve Eq. (\ref{map_shooting}) for $\mathbf{V}$, substitute $\mathbf{V}_{0}$ by $\mathbf{V}$ and solve again the seven initial value problems (\ref{ET3}) to find the new $\mathbf{X}_{0}$ and $\mathds{J}_{s0}$. This iterative process results in the following map

\begin{equation}\label{map_shooting_newton}
\mathbf{V}_{k} = \mathbf{V}_{k-1} + \mathds{J}_{sk}^{-1}(\mathbf{X}_{k}-\mathbf{X}_{k-1}),
\end{equation}

\noindent with $k=1,2,...$ and $\mathbf{X}_{k}=[\mathbf{x}_{{L}^{-}},\mathbf{x}_{{L}_{B}}]^{\intercal}$. If for each iteration $k$ the matrix $\mathds{J}_{sk}$ is well conditioned and the map (\ref{map_shooting_newton}) converges, these velocities solve the two-point boundary problem given by (\ref{ET3}) and (\ref{BC}) within the numerical error.

\par The condition number of the shooting matrix depends on the stability of the initial value problem \cite{Ascher_Petzold_1998,Ascher_1995} and with generic boundary segments one may not be able either to invert matrix $\mathds{J}_{sk}$ or to find a unique solution or any solution for map (\ref{map_shooting_newton}). As we show next, for small perturbations of circular orbits of large radii the boundary value problem is well-posed in a local subspace of boundary segments, matrix $\mathds{J}_{sk}$ is well conditioned and iteration (\ref{map_shooting_newton}) converges. In general, we expect the orbital velocities at points $\mathbf{O}^{+}$ and $\mathbf{L}^{-}$ to be different from the velocity on the boundary segments, and thus \emph{discontinuous}. 

\par In Theorem \ref{theorem1} we analyze the condition of matrix $\mathds{J}_{sk}$ and the convergence of map (\ref{map_shooting_newton}) for boundary-segments near segments of circular orbits of large radii.

\begin{theorem}\label{theorem1}
Let $\mathbf{x}_{i}^{S}(t) \in \mathbb{R}^3$ and $\mathbf{v}_{i}^{S}(t) \in \mathbb{R}^3$, denote the positions and velocities along a doubly circular orbit \cite{Schild_1963}, and $\mathbf{x}_{i}^{h}(t) \in \mathbb{R}^3$ be the boundary-segments for $i=1,2$. Assume the $\mathbf{x}_{i}^{h}(t)$ are \; $C^{2}$ and in a neighborhood of circular orbits with large radii, $r_{12}\gg 1$. Then the two-point boundary problem posed by ODE (\ref{ET3}) with boundary conditions (\ref{BC}) has a unique solution that can be found using the shooting method.
\end{theorem}

\noindent \textbf{Proof.} The positions are given by the integral 

\begin{equation}\label{int_theorem1}
\mathbf{X}(t_{{L}^{-}}) = \mathbf{X}(t_{O_{A}}) + \int_{t_{O_{A}}}^{t_{{L}^{-}}}\mathbf{V}(t)dt,
\end{equation}

\noindent and for boundary segments near circular orbits with large radii, the velocity $\mathbf{V}(t)$ is almost constant because the 
acceleration falls at the most as $1/r_{12}$, as can be found by inspecting (\ref{eq_WF4}) and (\ref{eq_Field2}). Notice that for near-circular boundary data of the shortest type the time span of the circular flight is so short that the trajectories are almost straight lines.

\par Using the above we have the approximation

\begin{equation}\label{app1_theorem1}
\mathbf{X}(t_{{L}^{-}}) \cong \mathbf{X}(t_{O_{A}}) + \mathbf{V}\Delta t.
\end{equation}

\noindent where $\Delta t = t_{{L}^{-}} - t_{O_{A}}$, $\mathbf{X}(t_{O_{A}}) = [\mathbf{x}_{1}^{S}(t_{O_{A}}),\mathbf{x}_{2}^{S}(t_{O_{A}})]^{\intercal}$, $\mathbf{X}(t_{{L}^{-}}) = [\mathbf{x}_{1}^{S}(t_{{L}^{-}}),\mathbf{x}_{2}^{S}(t_{{L}^{-}})]^{\intercal}$ and $\mathbf{V}$ is a constant vector. Approximating $\mathbf{V}$ by the initial circular velocity $\mathbf{V}_{0} = [\mathbf{v}_{1}^{S}(t_{O_{A}}),\mathbf{v}_{2}^{S}(t_{O_{A}})]^{\intercal}$ yields

\begin{equation}\label{app2_theorem1}
\mathbf{X}_{0} = \mathbf{X}( t_{O_{A}}) + \mathbf{V}_{0}\Delta t
\end{equation}

\noindent and Eqs. (\ref{app1_theorem1}) and (\ref{app2_theorem1}) yield

\begin{equation}\label{app3_theorem1}
\mathbf{X} = \mathbf{X}_{0} + \Delta t (\mathbf{V}-\mathbf{V}_{0}).
\end{equation}

\par Comparing (\ref{app3_theorem1}) with (\ref{map_shooting}) yields (\ref{map_shooting}) with $\mathds{J}_{s0} = \Delta t \mathds{1} $, where $\Delta t$ is the time span and $\mathds{1}$ the $6 \times 6$ identity matrix. Therefore matrix $\mathds{J}_{s0}$ has a well-conditioned inverse and the unique solution depends continuously on the boundary segments and initial velocities, i.e., a well-posed shooting method (\ref{map_shooting_newton}).\hspace{50.0mm}$\blacksquare$

\vspace{5mm}


\par When the continuous boundary-segments are only piecewise $C^2$ and have velocity discontinuities, one has to stop the integration of ODE (\ref{ET3}) at all breaking points to satisfy the Weierstrass-Erdmann corner conditions (\ref{WECCsec2}), as follows. 

\par For boundary segments with pre-specified jumps, the Weierstrass-Erdmann corner conditions (\ref{WECCsec2}) form an overdetermined system of $8$ equations for the $6$ velocities $\mathbf{v}_{1}^{r}$, $\mathbf{v}_{2}^{r}$ to continue the orbit on the right-hand-side of each discontinuity point. The former shows that for generic boundary segments the solution may not even exist. 

\par In order to describe boundary segments that \emph{do have} solutions with nontrivial discontinuous velocities, we include as variables two components of each right-velocity along each boundary segment, $\mathbf{v}_{2-}^{r}$, $\mathbf{v}_{1+}^{r}$, which have not been used until the corner point. Next we show that such augment of the set of variables must be made very carefully.
 
\par The above described augment generates an underdetermined nonlinear system having $8$ equations and $12$ variables,

\begin{equation}\label{WECCsec3}
  \left\{
    \begin{array}{ll}
    	\Delta\mathbf{P}_{1}(\mathbf{v}_{1}^{r},\mathbf{v}_{2}^{r},\mathbf{v}_{1+}^{r},\mathbf{v}_{2-}^{r}) = \mathbf{0}, \vspace{0.2cm}\\
	\Delta\mathbf{P}_{2}(\mathbf{v}_{1}^{r},\mathbf{v}_{2}^{r},\mathbf{v}_{1+}^{r},\mathbf{v}_{2-}^{r}) = \mathbf{0}, \vspace{0.2cm}\\
    	\Delta E_{1}(\mathbf{v}_{1}^{r},\mathbf{v}_{2}^{r},\mathbf{v}_{1+}^{r},\mathbf{v}_{2-}^{r}) = 0, \vspace{0.2cm}\\
	\Delta E_{2}(\mathbf{v}_{1}^{r},\mathbf{v}_{2}^{r},\mathbf{v}_{1+}^{r},\mathbf{v}_{2-}^{r}) = 0,
    \end{array}
  \right.
\end{equation}

\noindent where the superscripts ${}^{l}$, ${}^{r}$ denote the left-limit and right-limit of the velocities at the breaking points. Defining  the velocity jumps $\Delta\mathbf{v}_{i}\coloneqq \mathbf{v}_{i}^{r}-\mathbf{v}_{i}^{l}$, the vector of jumps $\bar{\mathbf{v}} \coloneqq {(\Delta\mathbf{v}_{1},\Delta\mathbf{v}_{2},\Delta\mathbf{v}_{1+},\Delta\mathbf{v}_{2-})}^{\intercal}$ and the local value of the continuous vector $\mathbf{d} \coloneqq {(\mathbf{P}_{1}^{l},\mathbf{P}_{2}^{l},E_{1}^{l},E_{2}^{l})}^{\intercal}$ we can rewrite Eq. (\ref{WECCsec3}) as

\begin{equation}
\mathbf{F}(\bar{\mathbf{v}}) = \mathbf{d}.
\end{equation}

\noindent Equations (\ref{momento_parcial}) and (\ref{energia_parcial}) are analytic functions of the velocities, thus yielding a locally convergent Taylor series for $\mathbf{F} (\bar{\mathbf{v}})$ in powers of the jump vector $\bar{\mathbf{v}} \coloneqq {(\Delta\mathbf{v}_{1},\Delta\mathbf{v}_{2},\Delta\mathbf{v}_{1+},\Delta\mathbf{v}_{2-})}^{\intercal}$. Expanding about some velocity $\bar{\mathbf{v}}_{0}$ yields 

\begin{equation}
\mathbf{F}(\bar{\mathbf{v}}_{0}) + \bar{\mathds{J}}_{0}(\bar{\mathbf{v}}_{0})(\bar{\mathbf{v}}-\bar{\mathbf{v}}_{0}) + \mathcal{O}(|\bar{\mathbf{v}}-\bar{\mathbf{v}}_{0}|^2)= \mathbf{d}.
\end{equation}


\noindent Defining $\bar{\mathbf{f}}_{0}\equiv\mathbf{d}-\mathbf{F}(\bar{\mathbf{v}}_{0})$, we obtain at lowest order a linear system of $8$ equations for $12$ variables,

\begin{equation}\label{Jacob12}
\bar{\mathds{J}}_{0}(\bar{\mathbf{v}}_{0})(\bar{\mathbf{v}}-\bar{\mathbf{v}}_{0}) = \bar{\mathbf{f}}_{0}.
\end{equation}

\par Next we analyze the linear system (\ref{Jacob12}) for near-circular boundary segments of large radii. The idea is to choose four of the twelve variables on the left-hand-side of (\ref{Jacob12}) as independent variables to be placed on the right-hand side of (\ref{Jacob12}). The system thus generated should have an invertible linear $8 \times 8 $ matrix with maximum row rank on the left-hand-side, yielding a unique solution for the eight ``slave variables" on the left-hand-side, i.e.,
  

\begin{equation}\label{Jacob81}
\mathds{J}_{0}(\mathbf{v}_{0})(\mathbf{v}-\mathbf{v}_{0}) = \mathbf{f}_{0}.
\end{equation}

\noindent Vector $\mathbf{f}_{0}\equiv\mathbf{d}-\mathbf{F}(\mathbf{v}_{0})$ on the right-hand-side of (\ref{Jacob81}) should depend on the four independent variables, and for a nontrivial discontinuity, $\mathbf{f}_{0}$ must be nonzero. 

\par We solve (\ref{Jacob12}) for $\mathbf{v}$ using the following iterative process. Starting from the solution $\mathbf{v}$ of linear system (\ref{Jacob81}), we replace $\mathbf{v}_{0}$ by $\mathbf{v}$ and recalculate $\mathds{J}_{0}(\mathbf{v}_{0})$ and $\mathbf{f}$ to find the next iterate $\mathbf{v}$ by (\ref{Jacob81}), thus generating the following map

\begin{equation}\label{MapWECC}
\mathbf{v}_{k+1} = \mathbf{v}_{k}+ \mathds{J}_{k}^{-1}(\mathbf{v}_{k})\mathbf{f}_{k},
\end{equation}

\noindent with $k=0,1,2,\ldots$ and $\mathbf{f}_{k}\equiv\mathbf{d}-\mathbf{F}(\mathbf{v}_{k})$.

\par Theorem \ref{theorem2} gives an example where linear system (\ref{Jacob81}) has a well conditioned matrix $\mathds{J}_{0}$ on the left-hand-side and a nonzero $\mathbf{f}_{0}$, thus determining a unique solution to (\ref{WECCsec3}) by iterating map (\ref{MapWECC}).

\begin{theorem}\label{theorem2} Let $( \mathbf{x}_{i}^{S}(t), \mathbf{v}_{i}^{S}(t) ) \in \mathbb{R}^3$ denote positions and velocities along a doubly circular orbit \cite{Schild_1963}, and $( \mathbf{x}_{i}^{h}(t), \mathbf{v}_{i}^{h}(t) ) \in \mathbb{R}^3$ be the $C^{2}$ boundary-segments in a neighbourhood of circular orbits with large radii, $r_{12}\gg 1$, for $i=1,2$ and with $\mathbf{v}_{2}^{h}(t)$ along the $\hat{x}$ axis and $\mathbf{v}_{1}^{h}(t)$along the $-\hat{x}$ axis. Then we can choose the $x$ and $z$ components of $\Delta \mathbf{v}_{1}^{h}(t)$ and $\Delta \mathbf{v}_{2}^{h}(t)$ as independent variables. The reduced linearized problem for the remaining (slave)  variables $\mathbf{v} \equiv {(\Delta\mathbf{v}_{1},\Delta\mathbf{v}_{2},\Delta {v}_{1+}^{y},\Delta {v}_{2-}^{y})}^{\intercal}$ is an inhomogeneous linear system given by (\ref{MapWECC}) with a well-conditioned matrix $\mathds{J}_{k}$, yielding a unique solution to Eq. (\ref{WECCsec3}).
\end{theorem}

\noindent \textbf{Proof.} For the above described near-circular orbits with a large inter-particle separation, $r_{12\pm} \equiv r \gg 1$, and low velocities, $v_{1\pm}^{r}=v_{2\pm}^{r} \equiv v^{r} \ll 1$, we have $\mathbf{n}_{12+}\approx\mathbf{e}_{y}$, $\mathbf{n}_{12-}\approx\mathbf{e}_{y}$, $\mathbf{v}_{2-}^{r}=v^{r}\mathbf{e}_{x}$, $\mathbf{n}_{21+}\approx \mathbf{n}_{21-}=-\mathbf{e}_{y}$, $v_{1+}^{r}=-v^{r}\mathbf{e}_{x}$. The linearized expansion (\ref{Jacob81}) evaluated in the limit of large radii where trajectories are approximated by straight lines near $\mathbf{v}_{0}=\mathbf{0}$ yields $\mathbf{f}_{0} = -\frac{1}{2r}{(\Delta {v}_{2-}^{x},0,\Delta {v}_{2-}^{z},\Delta {v}_{1+}^{x},0,\Delta {v}_{1+}^{z},0,0)}^{\intercal}$ and 

 \begin{equation}
	\mathds{J}_{0} \;=\,
	\left(\!\!\!\begin{array}{cccccccccccc}
  	m_{1} & 0 & 0 & \!\!-\frac{1}{2r}\!\! & 0 & 0 & 0 & 0 \vspace{0.2cm}\\
	0 & m_{1} & 0 & 0 & \!\!-\frac{1}{2r}\!\! & 0 & \!\!-\frac{1}{2r}\!\! & 0\vspace{0.2cm}\\
	0 & 0 & m_{1} & 0 & 0 & \!\!-\frac{1}{2r}\!\! & 0 & 0\vspace{0.2cm}\\
	\!\!-\frac{1}{2r}\!\! & 0 & 0 & m_{2} & 0 & 0 & 0 & 0 &\vspace{0.2cm}\\
	0 & \!\!-\frac{1}{2r}\!\! & 0 & 0 & m_{2} & 0 & 0 & \!\!-\frac{1}{2r}\!\!\vspace{0.2cm}\\
	0 & 0 & \!\!-\frac{1}{2r}\!\! & 0 & 0 & m_{2} & 0 & 0\vspace{0.2cm}\\
	-m_{1}v_{1}^{r} & 0 & 0 & 0 & \!\!\frac{1}{2r}\!\! & 0 & \!\!-\frac{1}{2r}\!\! & 0\vspace{0.2cm}\\
	0 & \!\!\frac{1}{2r}\!\! & 0 & m_{2}v_{2}^{r} & 0 & 0 & 0 & \!\!-\frac{1}{2r}\!\!\vspace{0.2cm}\\
	\end{array}\!\!\!\right).
\end{equation}

\noindent  Matrix $\mathds{J}_{0}$ is well-conditioned and one can calculate the next iterate $\mathbf{v}_{k+1}$ using map (\ref{MapWECC}) with $k=1$. For a continuous dependence on boundary data, velocity discontinuities in histories can be chosen arbitrarily in the subspace of independent discontinuity variables, $(\Delta {v}_{1+}^{x},\Delta {v}_{2-}^{x},\Delta {v}_{1+}^{z},\Delta {v}_{2-}^{z})$. Thus restricted, matrix $\mathds{J}_{k}$ is invertible and the unique solution for the slave variables $\mathbf{v} \equiv {(\Delta\mathbf{v}_{1},\Delta\mathbf{v}_{2},\Delta {v}_{1+}^{y},\Delta {v}_{2-}^{y})}^{\intercal}$ depends continuously on the boundary data and on $(\Delta {v}_{1+}^{x},\Delta {v}_{2-}^{x},\Delta {v}_{1+}^{z},\Delta {v}_{2-}^{z})$ and the unique solution to (\ref{WECCsec3}) is given by the fixed point of map (\ref{MapWECC}). \hspace{68.1mm}$\blacksquare$


\section{Numerical Experiments}\label{SecNum}

\par The family of circular orbits\cite{Schild_1963} with angular velocity $\omega$ and radius $r_{i}=v_{i}/\omega$, with $i=1,2$, can be parametrized by the retardation angle\cite{PIERB_2013} $\theta$. The light-cone time $\tau$ for light to travel the inter-particle distance is related to the constant retardation angle of the circular orbit by\cite{PIERB_2013} $\tau=\theta/\omega$. In the limit of small $\theta$, the circular radii are given by $r_i =1/(m_i \theta^2)$ for $i=1,2$ and the constant angular frequency is $\omega = m_1 m_2 \theta^3/(m_1 + m_2)$\cite{PIERB_2013}.

\par Our first numerical experiment uses boundary data given by a continuous perturbation of circular orbit's segments with $\theta=0.77$ for $m_{1}=1$ and $m_{2}=2$, henceforth boundary data (I). The perturbed boundary segments include a velocity discontinuity in the middle of the histories, i.e., are only piecewise $C^2$. The numerical solution is calculated with a shooting method that solves each initial value problem (\ref{ET3}) with a fourth-order Runge-Kutta method, as described in Section (\ref{SecODEandWECC}). When the integration reaches the breaking point the Runge-Kutta integrator is halted and we solve the Weierstrass-Erdmann corner conditions using the linear solution (\ref{Jacob81}) as initial guess for the function \textit{fsolve} of MatLab R2011a. 

\par
\begin{figure}[!htbp]\centering
\includegraphics[width=0.475\textwidth]{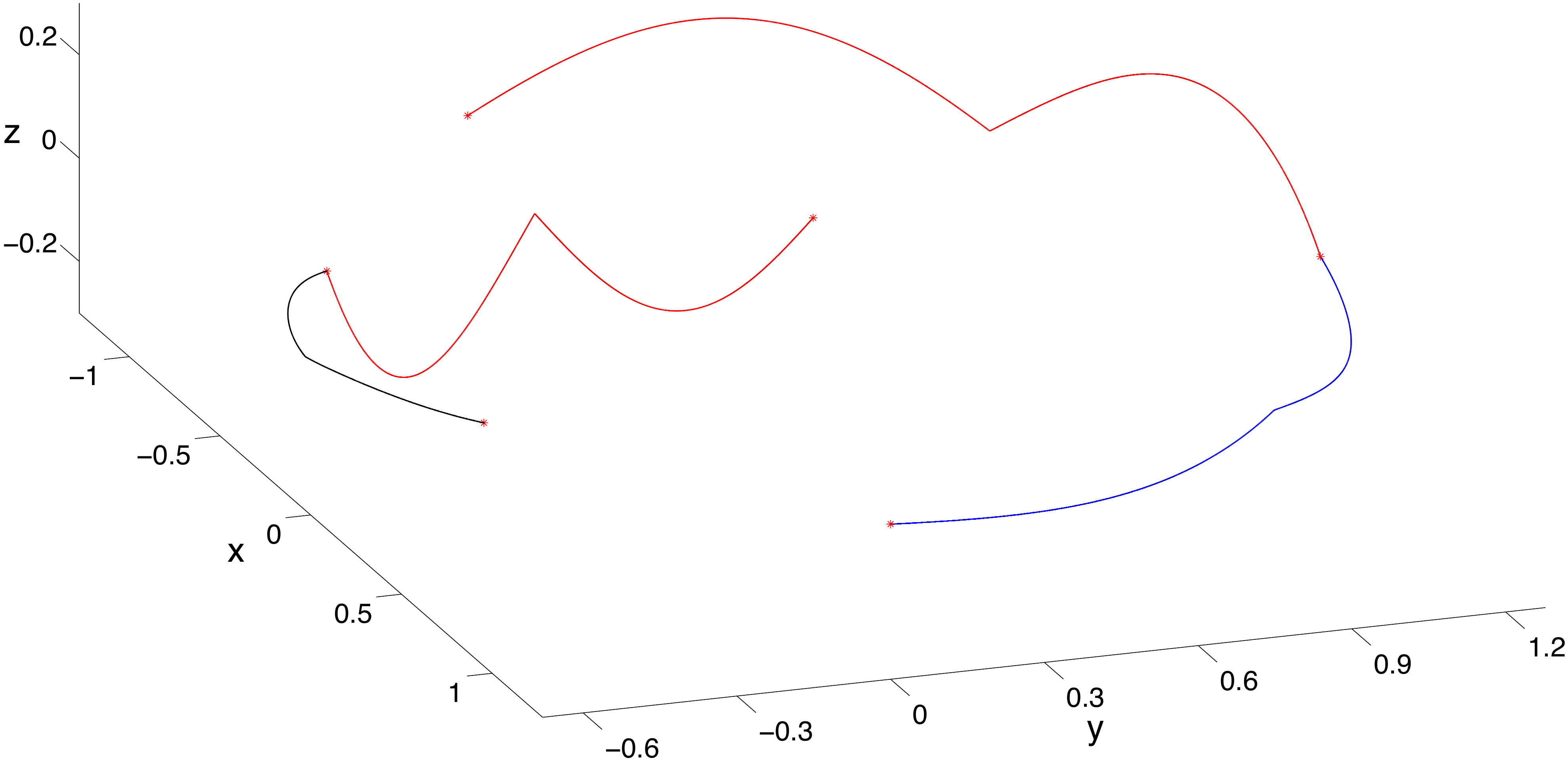}
\caption{\label{fig3}\footnotesize Trajectories for boundary data (I) having a single velocity-discontinuity point and in a neighbourhood of the a circular orbit with $\theta=0.77$, for $m_{1}=1$ and $m_{2}=2$. Trajectory of particle $1$ (solid blue line) and the future history segment $1$ (solid red line). Trajectory of particle $2$ (solid black line) and the past history segment $2$ (solid red line). Arbitrary units.}
\end{figure}

\begin{figure}[!htbp]\centering
\includegraphics[width=0.475\textwidth]{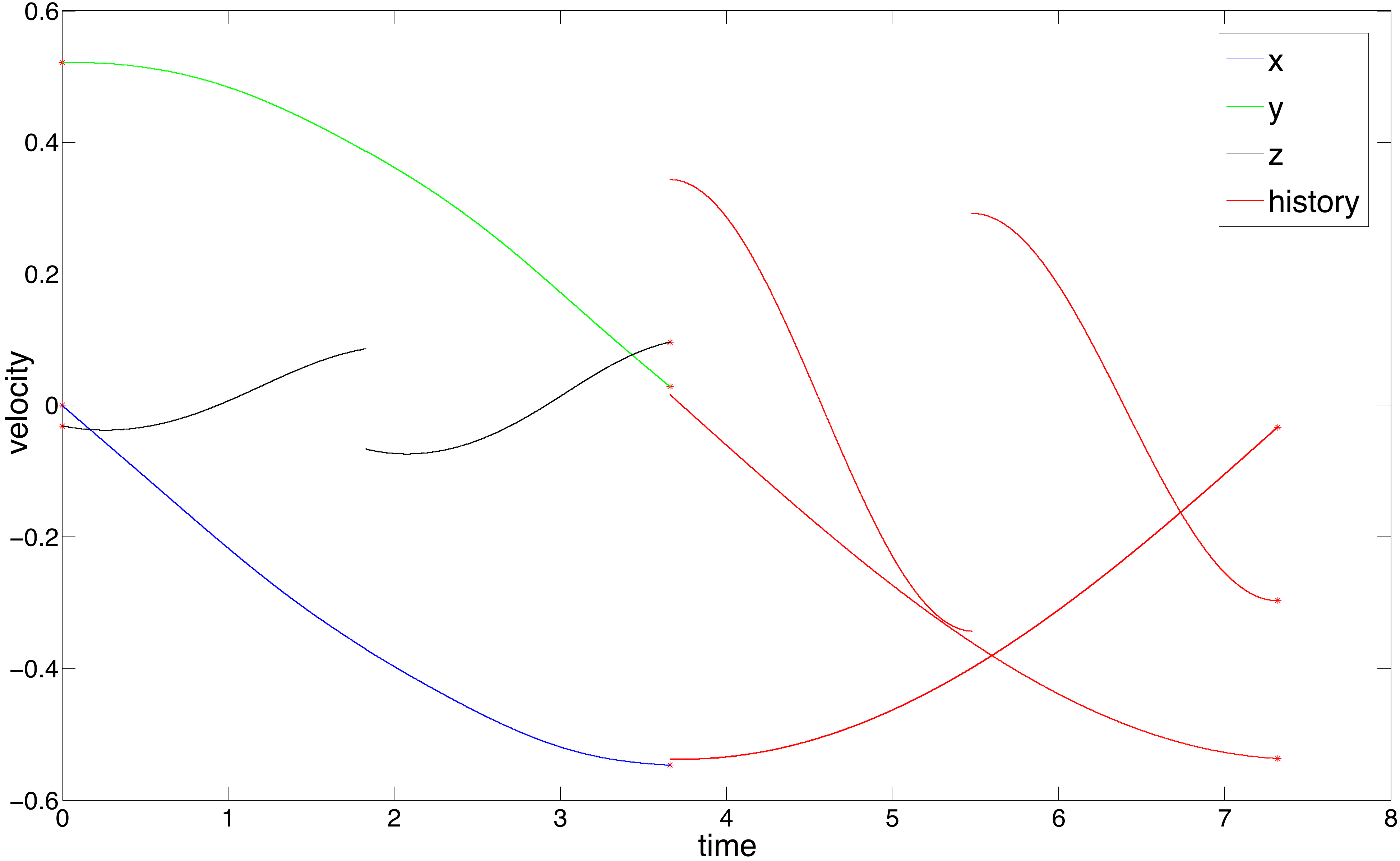}
\caption{\label{fig4}\footnotesize  Velocity of charge $1$ for boundary data (I), having one breaking point and given by perturbation in a neighbourhood of the circular orbit with $\theta=0.77$, $m_{1}=1$ and $m_{2}=2$. Arbitrary units.}
\end{figure}

\begin{figure}[!htbp]\centering
\includegraphics[width=0.475\textwidth]{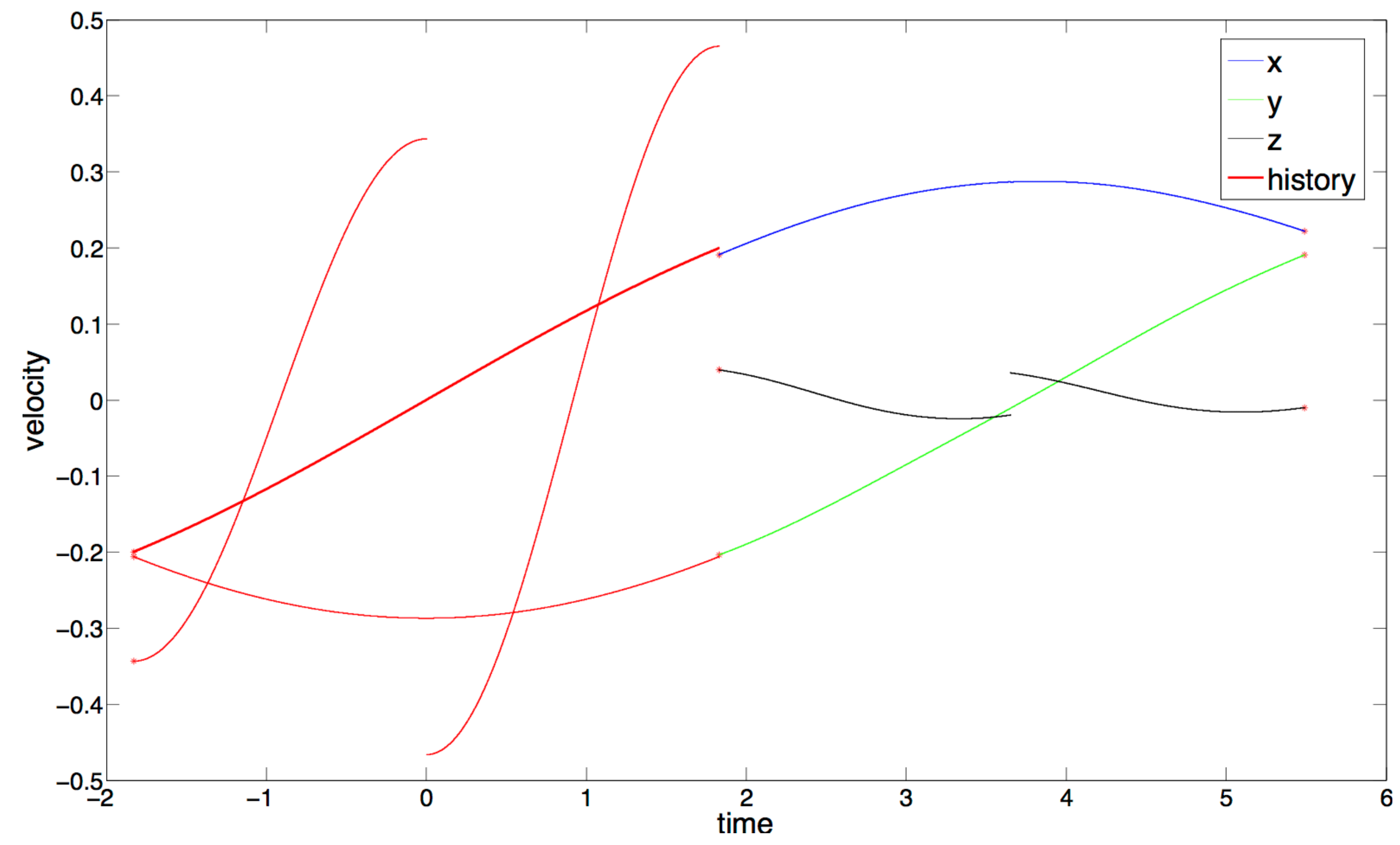}
\caption{\label{fig5}\footnotesize Velocity of charge $2$ for boundary data (I), having one breaking point and given by perturbation in a neighbourhood of a circular orbit with $\theta=0.77$, $m_{1}=1$ and $m_{2}=2$.  Arbitrary units.}
\end{figure}

 In Fig. \ref{fig3} we show the trajectories of the particles, history segments in red and numerically calculated trajectories in black and blue lines. In Fig. \ref{fig4} we show the components of the velocity of particle $1$ and Fig. \ref{fig5} shows the components of the velocity of particle $2$. Notice that the numerically calculated solutions have discontinuous velocities at points $\boldsymbol{O}^{+}$ and $\boldsymbol{L}^{-}$ (which is a generic feature of the shortest-lenght boundaries) and at one extra pair of points in light-cone and along each trajectory as caused by the breaking point in histories.
 

\par
\begin{figure}[!htbp]\centering
\includegraphics[width=0.475\textwidth]{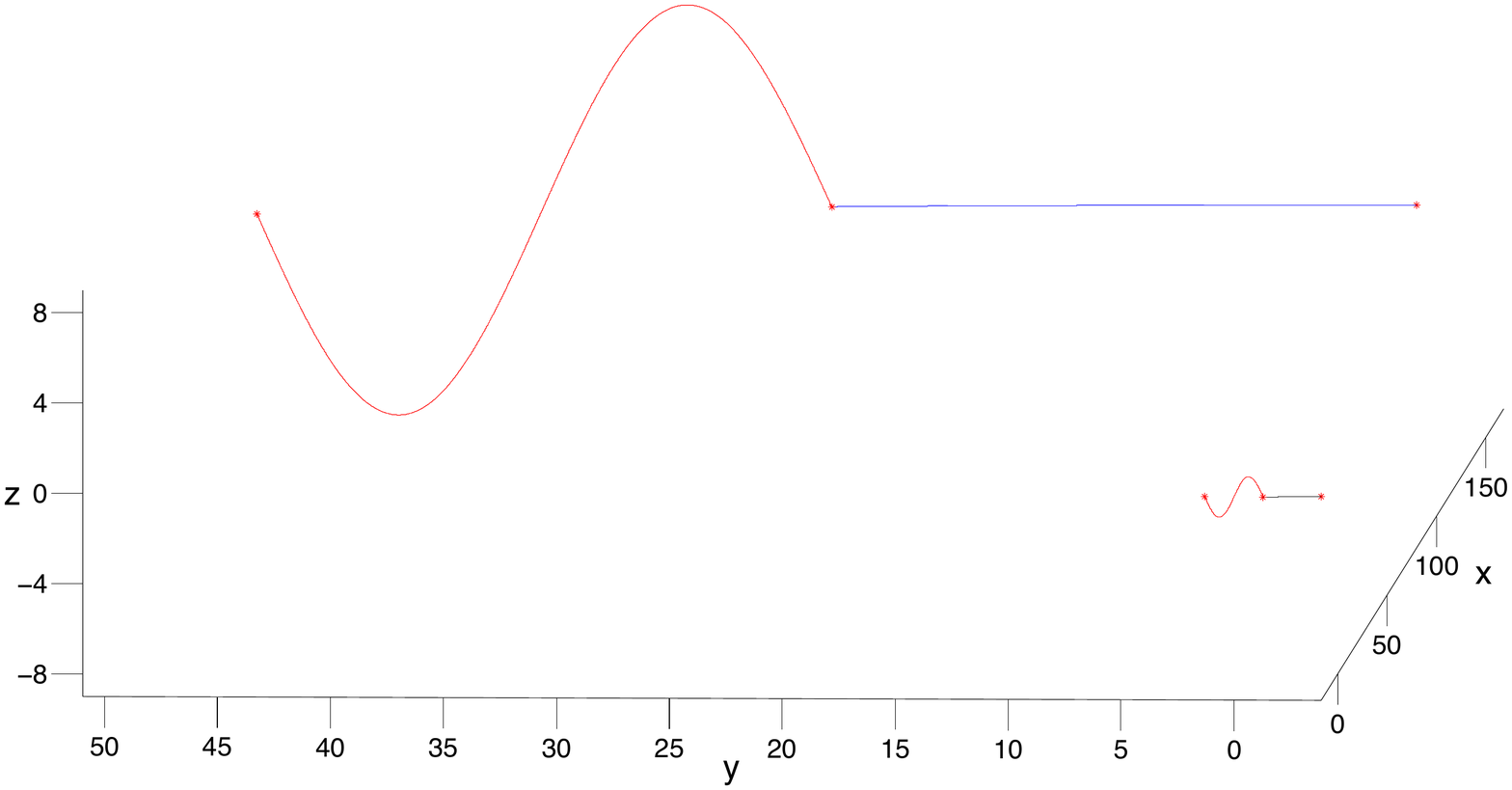}
\caption{\label{fig6}\footnotesize Trajectories for boundary data (II) given by a $C^2$ perturbation of a circular orbit with $\theta=0.077$, $m_{1}=1$ and $m_{2}=10$. Trajectory of particle $1$ (solid blue line) and the future history segment $1$ (solid red line). Trajectory of particle $2$ (solid black line) and the past history segment $2$ (solid red line). Arbitrary units.}
\end{figure}

\begin{figure}[!htbp]\centering
\includegraphics[width=0.475\textwidth]{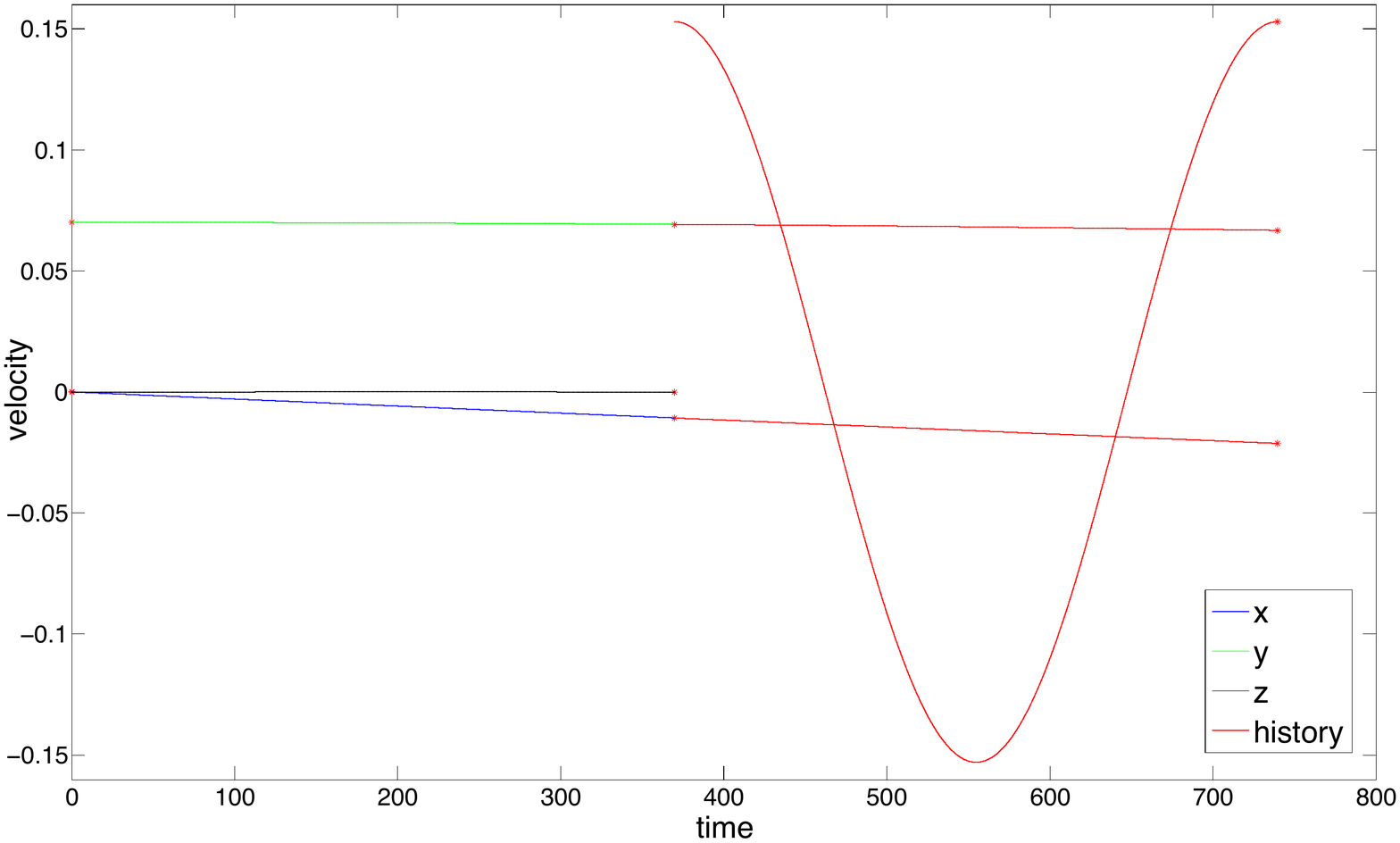}
\caption{\label{fig7}\footnotesize Numerically calculated velocity of charge $1$ for boundary data (II) given by a $C^2$ perturbation of a circular orbit with $\theta=0.077$, $m_{1}=1$ and $m_{2}=10$.  Arbitrary units. }
\end{figure}

\begin{figure}[!htbp]\centering
\includegraphics[width=0.475\textwidth]{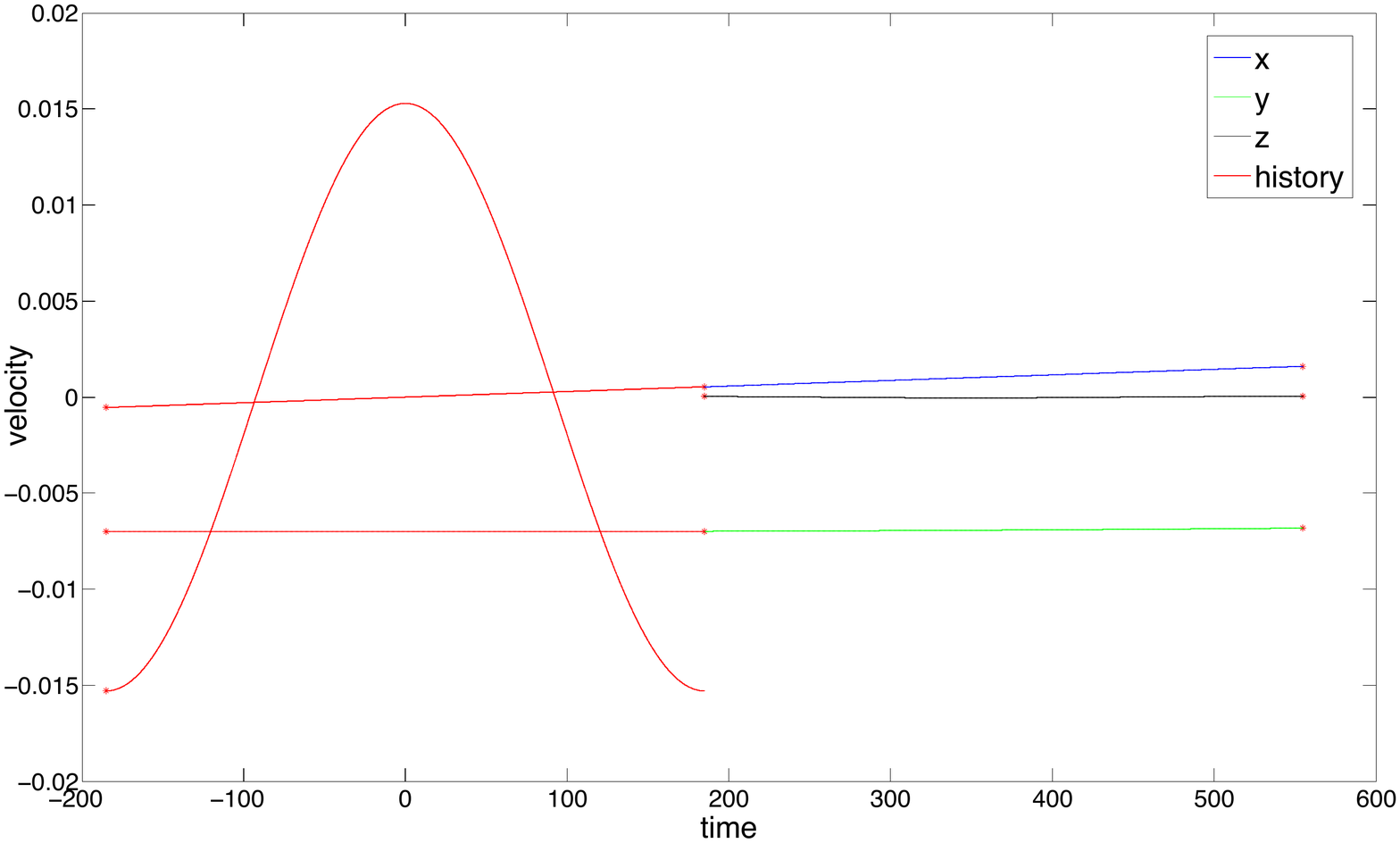}
\caption{\label{fig8}\footnotesize Velocity of charge $2$ for boundary data (II) given by a $C^2$ perturbation of a circular orbit with $\theta=0.077$, $m_{1}=1$ and $m_{2}=10$. Arbitrary units.}
\end{figure}

 In our second numerical experiment the boundary segments are given by a $C^2$ perturbation of circular orbit's segments with $\theta=0.077$ for $m_{1}=1$ and $m_{2}=10$, henceforth boundary data (II). The perturbed boundary segments have no velocity discontinuity in the history segments. The numerical solution in again calculated with a shooting method that solves each initial value problem (\ref{ET3}) with a fourth-order Runge-Kutta method, as described in Section (\ref{SecODEandWECC}).  In Fig. \ref{fig6} we show the trajectories of the particles, history segments in red and numerically calculated trajectories in black and blue lines. Last, Fig. \ref{fig7} shows the components of the velocity of particle $1$ and Fig. \ref{fig8} shows the components of the velocity of particle $2$. Notice that for our second experiment the numerically calculated solutions have velocity discontinuities only at points $\boldsymbol{O}^{+}$ and $\boldsymbol{L}^{-}$.  

\section{Discussions and Conclusion}\label{SecDiscConc}

\par We studied the variational boundary value problem of the electromagnetic two-body problem with shortest-length boundary-segments in a neighbourhood of circular orbits with large inter-particle separations, $r_{ij}\gg 1$. The Wheeler-Feynman equations reduce to a two-point boundary problem, (\ref{ET3}) and (\ref{BC}). For this case the initial value problem given by (\ref{ET3}) is well-posed and the solution is unique (Theorem \ref{theorem1}). We observed that the shooting method converged even for some perturbations of circular orbits with radii $r_{ij}\approx 0.5$ and having relativistic velocities $v_{i}\approx 0.9$, provided that $|\mathbf{n}_{ij\pm}\cdot\mathbf{v}_{j\pm}|\approx 0.3$.

\par The shooting problem uses up all the initial-velocity freedoms and the occurrence of discontinuous velocities at points $\boldsymbol{O}^{+}$ and $\boldsymbol{L}^{-}$ is expected even for $C^2$ perturbations of circular-orbit segments (i.e., without breaking points in histories). We have also shown existence of solutions with discontinuous velocities for near-circular boundary-segments having discontinuous velocities in histories. For boundary-segments with continuous velocities, trajectories may still have discontinuous velocities satisfying Eq. (\ref{WECCsec3}) for inter-particle separation in the nuclear magnitude $r_{12} \simeq 1/\sqrt{m_1 m_2}$, a case described by the algebraic-differential equation (\ref{eqAlg}). 

\par Situations where velocity discontinuities \emph{necessarily} occur are (a) one of the boundary-segments has discontinuous velocities, for example $\Delta\mathbf{v}_{1+}=\mathbf{0}$ and $\Delta\mathbf{v}_{2-}\neq\mathbf{0}$. In this case Eq. (\ref{WECCsec3}) necessarily predicts $\Delta\mathbf{v}_{1}\neq\mathbf{0}$ and $\Delta\mathbf{v}_{2}\neq\mathbf{0}$ and (b) both boundary-segments have discontinuous velocities, $\Delta\mathbf{v}_{1+}\neq\mathbf{0}$ and $\Delta\mathbf{v}_{2-}\neq\mathbf{0}$. For this case, Eq. (\ref{WECCsec3}) also predicts $\Delta\mathbf{v}_{1}\neq\mathbf{0}$ and/or $\Delta\mathbf{v}_{2}\neq\mathbf{0}$.


\begin{acknowledgments}
Daniel C\^{a}mara de Souza acknowledges the support of FAPESP doctoral scholarship 2010/16964-0 and Jayme De Luca acknowledges the partial support of FAPESP regular grant 2011/18343-6.
\end{acknowledgments}



\begin{thebibliography}{apsrmp4-1}

\bibitem{WheelerFeynman_1945} \textsc{J. A. Wheeler and R. P. Feynman}, \textnormal{Interaction with the Absorber as the Mechanism of Radiation}. \textit{Rev. Mod. Phys.} \textbf{17}, 157 (1945).

\bibitem{WheelerFeynman_1949} \textsc{J. A. Wheeler and R. P. Feynman}, \textnormal{Classical Electrodynamics in Terms of Direct Interparticle Action}. \textit{Rev. Mod. Phys.} \textbf{21}, 425 (1949).

\bibitem{Fokker_1929} \textsc{A. D. Fokker}, \textnormal{Ein invarianter Variationssatz f\"{u}r die Bewegung mehrerer elektrischer Massenteilchen}. \textit{Zeits. f. Physik} \textbf{58}, 386 (1929). 

\bibitem{Schwarzschild_1903} \textsc{K. Schwarzschild}, \textnormal{Zur Elektrodynamik. II. Die elementare elektrodynamische Kraft}. \textit{Gottinger Nachrichten} \textbf{128}, 132 (1903). 

\bibitem{Tetrode_1922} \textsc{H. Tetrode}, \textnormal{\"{u}ber den Wirkungszusammenhang der Welt. Eine Erweiterung der klassischen Dynamik}. \textit{Zeits. f. Physik} \textbf{10}, 317 (1922). 

\bibitem{WKB} \textsc{D. ter Haar}, \textnormal{The Old Quantum Theory}. Pergamon Press, New York (1967).

\bibitem{HansWKB} \textsc{C. M. Andersen and Hans C. Von Baeyer}, \textnormal{Circular Orbits in Classical Two-Body Systems}. \textit{Annals of Physics} \textbf{60}, 67-84 (1970).

\bibitem{Mehra} J. Mehra, \textsc{J. Mehra}, \textnormal{The Beat of a Different Drum: Life and Science of Richard Feynman}. Oxford University Press Inc., New York (1994).

\bibitem{JMP2009} \textsc{J. De Luca}, \textnormal{Variational Principle for the Wheeler-Feynman Electrodynamics}. \textit{J. Math. Phys.} \textbf{50}, 062701 (2009). 

\bibitem{minimizer} \textsc{J. De Luca}, \textnormal{Minimizers with discontinuous velocities for the electromagnetic variational method}. \textit{Phys.\ Rev.\ E} \textbf{82}, 026212 (2010). 

\bibitem{PIERB_2013} \textsc{J. De Luca}, \textnormal{Variational electrodynamics of atoms}. \textit{Progress In Electromagnetics Research B}. \textbf{53}, 147-186 (2013). 

\bibitem{Gelfand_1963} \textsc{I. M. Gelfand e S. V. Fomin}, \textnormal{Calculus of Variations}. Prentice-Hall, Inc., Englewood Cliffs (1963).

\bibitem{Bellen_Zennaro_2003} \textsc{A. Bellen and M. Zennaro}, \textnormal{Numerical Methods for Delay Differential Equations}. Oxford University Press, New York (2003).

\bibitem{Bellen_Guglielmi_2009} \textsc{A. Bellen and N. Guglielmi}, \textnormal{Solving neutral delay differential equations with state-dependent delays}. \textit{J. Comput. and App. Math.} \textbf{229}, 350-362 (2009).

\bibitem{Ascher_Petzold_1998} \textsc{U. M. Ascher and L. R. Petzold}, \textnormal{Computer Methods for Ordinary Differential Equations and Differential-Algebraic Equations}. SIAM, Philadelphia (1998).

\bibitem{Ascher_1995} \textsc{U. M. Ascher, R. M. M. Mattheij and R. D. Russel}, \textnormal{Numerical Solution of Boundary Value Problems for Ordinary Differential Equations}. SIAM, Englewood Cliffs (1995).

\bibitem{Schild_1963} \textsc{A. Schild}, \textnormal{Electromagnetic Two-Body Problem}. \textit{Phys. Rev.} \textbf{131}, 2762 (1963).

\bibitem{Schonberg_1946} \textsc{M. Sch\"{o}nberg}, \textnormal{Classical Theory of the Point Electron}. \textit{Phys. Rev.} \textbf{69}, 211 (1946).

\bibitem{Jackson_1999} \textsc{J. D. Jackson}, \textnormal{Classical Electrodynamics}. Third edition, John Wiley and Sons, New York (1999).

\bibitem{JCAMRev7} \textsc{J. De Luca, T. Humpries and S. B. Rodrigues}, \textnormal{Finite Element Boundary Value Integration of Wheeler-Feynman Electrodynamics}. \textit{J.\ Comput.\ Appl.\ Math.} \textbf{236}(13). 3319-3337 (2012).

\end{thebibliography}
\end{document}